# Local adsorption structure and bonding of porphine on Cu(111) before and after self-metalation

D. A. Duncan[1,2]*, P. Casado Aguilar[1], M. Paszkiewicz[1], K. Diller[1,3], F. Bondino[4], E. Magnano[4,5], F. Klappenberger[1], I. Píš[4,6], A. Rubio[7,8,9], J. V. Barth[1], A. Pérez Paz[7,10]*, F. Allegretti[1]*

[1] Technical University of Munich, Department of Physics E20, 85748 Garching, Germany.

[2] Diamond Light Source, Harwell Science and Innovation Campus, Didcot, OX11 0DE, United Kingdom.

[3] Technical University of Munich, Department of Chemistry, 85748 Garching, Germany.

[4] IOM-CNR, Laboratorio TASC, S.S. 14-km 163.5, 34149 Basovizza, Trieste, Italy.

[5] Department of Physics, University of Johannesburg, PO Box 524, Auckland Park, 2006, Johannesburg, South Africa

[6] Elettra-Sincrotrone Trieste S.C.p.A., S.S. 14-km 163.5, 34149 Basovizza, Trieste, Italy

[7] Nano-Bio Spectroscopy Group and ETSF, Universidad del País Vasco, 20018 San Sebastián, Spain

[8] Max Planck Institute for the Structure and Dynamics of Matter, Luruper Chaussee 149, 22761 Hamburg, Germany

[9] Center for Free-Electron Laser Science & Department of Physics, University of Hamburg, Luruper Chaussee 149, 22761 Hamburg, Germany

[10] School of Physical Sciences and Nanotechnology, Yachay Tech University, Urcuquí 100119, Ecuador

**Abstract**

*We have experimentally determined the lateral registry and geometric structure of free-base porphine (2H-P) and copper-metalated porphine (Cu-P) adsorbed on Cu(111), by means of energy-scanned photoelectron diffraction (PhD), and compared the experimental results to density functional theory (DFT) calculations that included van der Waals corrections within the Tkatchenko-Scheffler approach. Both 2H-P and Cu-P adsorb with their center above a surface bridge site. Consistency is obtained between the experimental and DFT-predicted structural models, with a characteristic change in the corrugation of the four N atoms of the molecule's macrocycle following metalation. Interestingly, comparison with previously published data for cobalt porphine adsorbed on the same surface evidences a distinct increase in the average height of the N atoms above the surface through the series 2H-P, Cu-P, cobalt porphine. Such an increase strikingly anti-correlates the DFT-predicted adsorption strength, with 2H-P having the smallest adsorption height despite the weakest calculated adsorption energy. In addition, our findings suggest that for these macrocyclic compounds, substrate-to-molecule charge transfer and adsorption strength may not be univocally correlated.*

1. Introduction

When discussing the strength of interaction between the atomic or molecular constituents of matter that are brought into contact with each other, various "rules of thumb" have been developed in physical and material chemistry. A simple exemplary concept, as frequently invoked for interfacial systems, proclaims that shorter bond lengths correlate with stronger interactions[1]. This idea is borrowed from molecular and solid-state chemistry, where the empirically established correlation of bond strengths and bond lengths[2-5] is often an implicit assumption in the interpretation of molecular and crystallographic structures. However, it is known in the scientific literature that this is an oversimplified view (e.g. Refs. 6-9) and, especially for dative bonds, the bond length can be a poor indicator of interaction strengths[10]. Accordingly, it has been pointed out that the adsorption height and interaction strength at adsorbate/metal interfaces do not necessarily correlate[11-13], as we will illustrate by way of an example herein. Along a similar vein, regarding such interfacial hetero-systems, it was hypothesized that the level of charge transfer between the adsorbate and the metal substrate can be interpreted as a measure of the interaction strength (e.g. Refs. 12-14), since the latter depends dominantly on the charge transfer in the simplified interpretation of polar, covalent and dative bonding. As we will emphasize in this work, however, also this second rule of thumb may be misleading when considering the anchoring and interaction of macrocyclic species and complexes on metallic substrates. Therefore, at least for these interfacial systems both the correlation between bond length and interaction strength and that between charge transfer and interaction strength should be taken with a "grain of salt".

In general, the study of metal organic complexes anchored on well-defined solid surfaces has raised significant interest over the last decades due to their diverse chemical reactivity[15-17], their intriguing electronic properties[18-20], and the ability to self-assemble into organized supramolecular networks and mediate charge-transfer processes or catalytic conversions[21-27]. Surface science experiments with tetrapyrrole molecules[22, 23, 28], in particular, have been a prominent field of research due to the high stability, the ease of deposition and the rich surface chemistry afforded on metal substrates. For example, they have been observed to engage in short chain oligomerization resulting in bandgap narrowing[29], and to bind to the edges of graphene to allow the incorporation of metal atoms into this "wonder material"[30]. Moreover, they have been used to probe the fundamental principles underlying the interactions between metallic substrates and metal-organic complexes[31, 32], and the acceptor-donor relationship in organic semiconductors[33]. One of the most notable characteristics of the tetrapyrrole molecules is that they can both coordinate a variety of elements into their central cavity and be substituted at terminal positions, which is useful to engineer a multitude of functionalities relevant for technological applications based on this class of molecules.

The most conspicuous tetrapyrrole molecules are porphyrins, of which the simplest example is porphine. Specifically, porphine consists solely of the four (modified) pyrrole groups linked by four methine (=CH-) bridges and, due to its prototypical character, it has recently been the focus of several fundamental studies[34-41]. The non-metalated, so-called "free-base" species (denoted here by 2H-P) is shown schematically in Fig. 1a. 2H-P and free-base porphyrins in general are often used as precursors for experiments on this family of molecules, as they can be easily metalated *in situ* with a wide variety of metal atoms[28, 38, 42]. Free-base porphyrins exhibit two peaks in their N 1$s$ X-ray photoelectron spectroscopy (XPS) data[28, 38, 42-45], whereas almost all metalated porphyrins exhibit only a single peak related to the macrocyclic nitrogen species[28, 38, 42-45]. The separation of the N 1$s$ XP spectra into two features, for 2H-P, is due to the chemical shift between the amino (N-H, higher binding energy) and imino (=N-, lower binding energy) moieties, which, as shown in Fig. 1a, each comprise two N atoms on opposite sides of the porphine cavity. For the metalated species, only a single peak is observed in the N 1$s$ XP spectra due to the chemical equivalency of the four macrocyclic N atoms interacting with the metal center, as shown in Fig. 1b.

Free-base porphyrins can "self-metalate" on several surfaces by capturing a substrate metal atom and incorporating it at the center of the macrocycle[38, 42, 46-49]. This has most famously been observed on the Cu(111) surface, where many of these self-metalation experiments were performed[38, 46, 50-55]. Specifically, the self-metalation of tetraphenyl porphyrin (TPP), in which each methine bridge is also bound to a phenyl ring, has been studied by Bürker *et al.* employing normal incidence X-ray standing waves (NIXSW)[53]. The adsorption height of the two distinct N species was found to be 2.02 ± 0.08 Å (imino) and 2.23 ± 0.05 (amino) Å above the Cu(111) surface before metalation, and 2.25 ± 0.02 Å, averaged over the four N atoms, after metalation with copper. Conversely, a similar NIXSW study into the adsorption of cobalt porphine, which was metalated prior to deposition onto Cu(111), was performed by Schwarz *et al.* and reports a larger mean adsorption height of the N atoms at 2.33 ± 0.06 Å[41]. These findings, and specifically the measured height discrepancy between cobalt porphine and copper TPP, could suggest that either (1) the substituted phenyl rings on the periphery of the TPP molecule result in a lower adsorption height or (2) the nature of the central metal atom has a significant effect on adsorption heights. The first possibility can be ruled out, however, because the phenyl groups are expected – if anything – to cause a lift of the entire porphyrin to alleviate the steric hindrance with the substrate. Additionally, the latter study of cobalt porphine directly demonstrated that the molecule adsorbs with its center located above a surface bridge site[41], whereas no experimental determination of the adsorption site is reported for copper TPP (Cu-TPP). This leaves open a third possibility that the usual condition of site-specific interaction between the iminic nitrogen atoms and the substrate may not be fulfilled upon metalation[53], and thus varying adsorption sites are responsible for the measured average height difference.

Within this background, we present herein an energy-scanned photoelectron diffraction (PhD)[56] study into the local adsorption site of both 2H-P and its self-metalated product, copper porphine (Cu-P – shown schematically in Fig. 1b) on the Cu(111) surface. The PhD technique is based on core-level XPS and exploits the interference between the component of the photoelectron wave-field that is emitted directly towards the detector, and those components that are emitted towards the underlying substrate and back-scattered in the direction of the detector[56, 57]. By varying the incident photon energy, and thus the photoelectron kinetic energy and wavelength, these different components are brought in and out of phase, resulting in a modulation of the photoelectron intensity at the detector. In a single scattering approximation, these modulations are a hologram image of the underlying substrate[58], with the directly emitted component acting as the reference beam and the scattered components as the object beam. However, due to the highly interacting nature of the low-energy electrons used in such studies, there is usually a significant contribution to the modulations from higher order scattering events, necessitating comparison to theoretical multiple scattering calculations. By measuring N 1$s$ PhD spectra of both 2H-P and Cu-P and simulating the diffraction modulations within an iterative "*trial-and-error*" fitting procedure, we are able to tease out the differences in the N adsorption height before and after metalation, allowing us to better understand the differences between the studies of Bürker *et al.*[53] and Schwarz *et al.*[41]. Moreover, with the aid of density functional theory (DFT) calculations, we extract trends in the adsorption energy, charge transfer and adsorption height of the closely related 2H-P, Cu-P and cobalt porphine molecules, and thus systematically compare how they interact with the Cu(111) surface.

## 2. Experimental and Computational Details

### 2.1 Sample preparation and experimental measurements

The soft X-ray photoelectron spectroscopy (SXPS) and PhD measurements, complemented by low-energy electron diffraction (LEED) experiments, were performed in the end station of the BACH undulator beam line at the Elettra Sincrotrone Trieste (Italy) with a base pressure of $3 \times 10^{-10}$ mbar. This end station comprises a preparation chamber, equipped with standard facilities for sample cleaning and high temperature annealing, and an analysis chamber equipped with a VG-Scienta R3000 hemispherical electron analyzer and LEED optics from OCI Vacuum Microengineering Inc.. The analyzer has vertical entrance slits and is mounted in the plane of the photon polarization (linear horizontal) at an angle of 60° with respect to the incident light[59]. A Cu(111) single crystal (Surface Preparation Laboratory, SPL) was prepared by repeated cycles of sputtering and annealing (670 K). Sample cleanliness and orientation were monitored by SXPS and LEED. 2H-P (Frontier Scientific, >95% purity) was evaporated onto the surface kept at room temperature (RT) using a home built Knudsen cell and an evaporation temperature of 470 K. This corresponded to an evaporation flux, at the sample, of under $1.5 \times 10^{16}$ molecules·m$^{-2}$·s$^{-1}$, based on the resulting LEED pattern. The 2H-P powder was thoroughly outgassed in the Knudsen cell by heating it to a temperature of 430 K and holding it there for ~1 hour. The cleanliness and chemical identity of the deposited molecular layer were also assessed by SXPS (not shown), and the coverage was monitored by measuring N $K$-edge near-edge X-ray absorption fine structure (NEXAFS) spectra (not shown). Both resulted in identical spectroscopic signatures as in the work of Diller et al.[38] for a full monolayer. Metalation was achieved upon annealing the Cu(111) sample to ~430 K and monitored by recording the N 1$s$ SXP spectra until only a single peak was observed.

N 1$s$ PhD scans (hv = 490 – 750 eV) were acquired over a range of emission angles and two different crystal orientations at RT. As the permanent end station in BACH did not allow for azimuthal rotation at the time of the experiment, the different crystal orientations were obtained by removing the crystal from ultra-high vacuum (UHV), physically remounting it with the desired azimuthal orientation, returning the crystal to UHV, cleaning and re-preparing the layer. A full set of PhD scans was acquired for the 2H-P in the $[1\bar{1}0]$ and $[11\bar{2}]$ directions (cf. Fig. 1c). As the two imino (amino) N atoms are related to each other across the mirror plane(s) of the molecule, and the molecule is expected, from previous DFT calculations[40] and STM measurements[38], to adsorb with one of the molecular mirror planes coincident with the surface mirror plane, the $[\bar{1}\bar{1}2]$ and $[11\bar{2}]$ directions were expected to be almost identical. Therefore, as time at the synchrotron is intrinsically limited, data acquisition along the $[\bar{1}\bar{1}2]$ direction was not pursued. Furthermore, due to time constraints, a full data set for the N 1$s$ PhD of Cu-P was only acquired for a single azimuthal direction, the $[11\bar{2}]$ direction, together with a normal emission (NE) measurement along the $[1\bar{1}0]$ direction.

Note that, since the acquisition time for PhD measurements is lengthy (~2-3 hours per experimental geometry), beam damage frequently represents an issue with this technique (e.g. Ref. 60). To assess this, N 1$s$ and C 1$s$ SXP spectra were acquired before and after each PhD scan and exhibited no evidence of radiation damage. Moreover, due to similar concerns of possible damage caused by low-energy electrons, LEED patterns were only acquired at the end of the study presented here; as such they were not acquired for 2H-P/Cu(111), though we do not expect the lateral ordering of the 2H-P molecules on Cu(111) to differ dramatically.

**2.2 Photoelectron diffraction analysis**

The PhD data, normalized to the reference current, were analyzed following the standard procedure for data reduction outlined in Ref. 56, where the secondary electron background is subtracted using a template background and the peaks are deconvoluted and fitted with Gaussian lineshapes. The integrated area of these Gaussian lineshapes are then plotted against their varying kinetic energy, fitted with a stiff spline which is subtracted from, and then used to normalize, the integrated area to obtain the (dimensionless) modulation in the photoemission intensity that is entirely due to diffractive effects, $\chi_{exp}$[56, 57]. The resulting modulations were then modelled ($\chi_{the}$) using the multiple scattering codes developed by Fritzsche[61-65] exploiting an implementation of a particle swarm optimization global search algorithm[66]. The agreement between theory and experiment, and thus the fitting factor for the global search, was assessed quantitatively *via* a Pendry-like R-factor[67] calculated by:

$$R_{fac} = \frac{\sum_E (\chi_{(E)exp} - \chi_{(E)the})^2}{\sum_E (\chi_{(E)exp}^2 + \chi_{(E)the}^2)}, \qquad (1)$$

where the sums are performed over all experimentally collected energy data points and all modulations acquired in different photoelectron emission geometries. The R-factor is defined such that it is 0 when the experiment and theory are in perfect agreement, 1 when not correlated, and 2 when anti-correlated. For a suitably large dataset of modulations that can be fitted with at most few unique structures, the best R-factors found tend to be in the order of <0.2 for modulations of greater than 40% and in the order of <0.4 for modulations weaker than 20%, which reflects the inherently poorer signal to noise ratio of these data. Similarly, the comparatively low coverage on the surface (~0.15 N atoms per 1 Cu atom) of the molecules will also result in a worse signal to noise ratio, which will also impact on the best achievable R-factor. The uncertainties in structural parameters are estimated by calculating the R-factor as a function of a given parameter, and determining when the R-factor exceeds the variance[68]. This difference, between the value of the parameter at the minimum R-factor ($R_{min}$) and where the R-factor just exceeds the variance ($R_{min}$ + var($R_{min}$)), is considered to be a single standard deviation (cf. Ref. 56 for the definition of var($R_{min}$)). Note that in the modelling of the adsorption site of both molecules, due to carbon being a weak scatterer, PhD measurements are largely insensitive to its position. Therefore, although the carbon atoms were included in the calculations, their positions were not explicitly fitted. Instead, the pyrrolic carbons were assumed to be co-planar with their aminic/iminic N atom, and the methine bridges to have an adsorption height half-way between their neighboring pyrrole rings. The H atoms were omitted, as these are extremely weak scatterers. Finally, note that for the final set of refined multiple scattering calculations, a subset of all experimental PhD spectra was selected. These are in general the spectra that exhibit the strongest modulations, which are the most statistically reliable, but it is also helpful to include as wide a range of emission geometries as possible. Such subsets for 2H-P and Cu-P are shown in Figs. 3 and 4, respectively.

**2.3 Computational details**

Density functional theory (DFT) slab calculations were performed using the projector augmented wave pseudo-potential method[69] as implemented in the VASP code (version 5.4.1)[70-72]. The Perdew, Burke and Ernzerhof (PBE) exchange-correlation functional was employed in all calculations[73]. These included van der Waals (vdW) corrections *via* the Tkatchenko-Scheffler approach[74]. The convergence threshold of the electronic cycle was set to $10^{-5}$ eV, and a Gaussian smearing of 0.1 eV was used. All geometry optimizations used a converged kinetic energy cut-off of 450 eV. Calculations with higher cut-off (500 eV) show negligible variations in the results. A converged 5×5×1 Monkhorst-Pack *k*-point mesh and "PREC=accurate" settings in VASP were employed. The Cu(111) surface was modelled with the coordinates derived from a typical PBE lattice constant of 3.63 Å [40, 75] in order to minimize the stress in the system. The low adsorbate coverage limit was investigated *via* an 8×8×4 Cu(111) slab (256 atoms). The computational cell was [[*L*, 0.0, 0.0], [*L* × cos(π/3), *L* × sin(π/3), 0.0], [0.0, 0.0, 30.0 Å]], where $L = 4\sqrt{2}a$ = 20.534 Å is the cell size (and also the separation between porphine centers) and *a* = 3.63 Å the above lattice constant. An 18 Å vacuum layer and dipole corrections were used to decouple the periodic images along the normal *z* direction. The 2H-P and Cu-P molecules were initially positioned with opposite N atoms along the bridge position of Cu(111)[76], similar to the adsorption mode of 2H-P/Cu(111) reported by Müller *et al.*[40], which places the two amino groups N-H along the long bridge site direction (cf. Fig. 1c). The adsorbate and the two uppermost Cu layers were relaxed until all ionic forces were below 0.025 eV/Å. We computed adsorption energies as the total DFT energy of the relaxed system minus the energy of its separate relaxed components (clean Cu(111) slab and porphine). Finally, the charge transfer was computed *via* the Bader analysis code[77-79]. Partial atomic charges are not quantum-mechanical observables and therefore the partition of electronic charge across atoms in the system is somewhat arbitrary and not uniquely defined in any *ab initio* calculation. Several partial charge schemes have been proposed (Bader, Mulliken, Löwdin, Hirshfeld, etc.), each giving varying charge transfer results depending on the specific system. Here, the Bader partitioning scheme was adopted, so as to enable direct comparison with previous calculations[41, 80].

## 3. Results

### 3.1 DFT study – predicted local structure

Our dispersion corrected DFT calculations using the 8×8×4 Cu(111) slab yielded an adsorption energy of -4.032 eV for 2H-P and -4.257 eV for Cu-P, respectively. Dispersion corrections are essential for both systems, and without them molecular adsorption was found to be unstable (i.e. positive adsorption energy or repulsive). 2H-P and Cu-P were predicted to adsorb with their center of mass above a bridge site (shown schematically in Fig. 1c), in agreement with previous findings for 2H-P[40] and cobalt porphine on Cu(111)[41].

For 2H-P, the amino groups (N-H) are found to sit astride the long bridge site direction (Fig. 1c) with the N atoms in a direct atop site, whereas the imino groups (=N-) are astride the short bridge site (Fig. 1c), 0.68 Å off atop. The bond distances with respect to the nearest neighbor Cu atom of the amino and imino species were 2.43 and 2.15 Å, respectively. The corresponding adsorption heights relative to the average Cu(111) top layer were 2.33 Å (amino) and 2.09 Å (imino), with a difference of 0.24 Å in height between the two types of N species.

Upon metalation, the N atoms residing along the long bridge direction, referred to as $N_{long}$, move closer to the substrate (height of 2.10 Å) and slightly off atop (0.12 Å). The N atoms along the short bridge ($N_{short}$) axis move, with respect to 2H-P, more atop (0.59 Å off atop) and further from the substrate (height of 2.21 Å). The N atoms along the short bridge axis are thus 0.11 Å higher than those along the long bridge. The Cu metal center is predicted to lie 0.09 Å below the mean adsorption height of the four N atoms, basically co-planar to the $N_{long}$ species. Further structural parameters are given in Table 1 (cf. also Fig. 1c, d). For cobalt porphine, DFT predicts two nearly isoenergetic adsorption modes that differ by a 90° rotation[41]. However, due to the more planar nature of the Cu-P molecule, energy minimization of a 90°-rotated Cu-P molecule resulted in essentially the same structure with negligible differences in the macrocycle (pertaining entries in Tables 1 and 2, and the Supplementary Material for Cu-P are actually averages between these two Cu-P conformations).

Our DFT calculations predict that both molecules receive a charge transfer over 1e$^-$ from the substrate, specifically 1.08 e$^-$ for 2H-P and 1.26 e$^-$ for Cu-P. This is in contrast to the reduced charge transfer (0.66 e$^-$) into cobalt porphine[41]. Note that the two N species in cobalt porphine were predicted to adsorb at a height of 2.37 Å and 2.48 Å in the short / long bridge direction, significantly higher than that predicted for 2H-P and Cu-P. The DFT-predicted mean heights of the N atoms relative to the mean top Cu(111) layer in the progression 2H-P, Cu-P, cobalt porphine are 2.21, 2.16 and 2.42 Å, respectively. In addition, Cu-P adsorbs essentially flat (tilt of < 5°) on the substrate, whereas the 2H-P macrocycle is predicted to exhibit a 7° upwards tilt of the pyrrolic groups (see inset of Fig. 5b,d) along both the <211> and <110> directions. In contrast, cobalt porphine features a more distorted macrocycle of asymmetric saddle shape. Of the two energetically degenerate models predicted to exist, with the pyrrole rings aligned along either the <211> or <110> directions significantly bent away from the substrate[41], the STM measurements suggested that it is the <211> orientation that is favored on the surface, however prior STM measurements of Cu-P on Cu(111)[38] showed a similar motif as cobalt porphine, with an identical orientation. Therefore, it is most likely that the contrast observed in STM for Cu-P (and probably cobalt porphine) is electronic, rather than topographical.

**3.2 LEED – lateral organization of Cu-P layer on Cu(111)**

The measured LEED patterns, for Cu-P on Cu(111), are shown in Fig. 2. These patterns exhibit three clear sets of fractional order spots that share the point group symmetry of the substrate, but are rotated 30° with respect to it. Therefore, this pattern is likely to correspond to a $(n\sqrt{3} \times n\sqrt{3})R30°$ overlayer. Fitting this pattern using the software LEEDpat[81] indicates that it is the $n = 3$ superstructure (see Fig. 2, right), with two sets of the fractional order spots missing. The origin of the missing spots is almost certainly that they are too dim to see, considering that the visible spots are all clustered around the substrate integer order spots. Specifically, a primitive $(3\sqrt{3} \times 3\sqrt{3})R30°$ overlayer is not only commensurate with the underlying substrate (see Fig. 5e), but also features a unique adsorption site for all Cu-P molecules with $3\sqrt{3} \times \frac{a}{\sqrt{2}} = 13.3$ Å separation between their centers. This distance is comparable to the average intermolecular separation of 2H-P on Ag(111)[37] and Ni porphine on the same surface[35] and would suggest a minimum intramolecular atom-atom distance of ~3 Å, as the diameter of the molecule is approximately 10 Å.

**3.3 PhD analysis of 2H-P on Cu(111)**

As the 2H-P molecule exhibits two peaks in its N 1s XP spectra, the amount of individual modulations potentially available for this system is higher than for the metalated species. The measurement geometries displaying the largest modulations are shown in Fig. 3. Strong, long period modulations can be observed in the normal emission (NE) data, suggesting an atop or near atop site, with the NE modulation for the amino (N-H) species being more intense than that of the imino (=N-) species. This indicates that the amino species are presumably situated more directly above substrate atoms than the imino species. The best fitting structure (shown in Fig. 3), resulting from the global search optimization, has an R-factor of 0.34 and indicates a difference in height between the amino and imino species of 0.12 ± 0.06 Å, with the latter being closer to the substrate. The corresponding difference in the N to Cu bond length is 0.16 ± 0.07 Å (cf. Table 1). The amino species was found to be astride the long bridge direction and the imino species astride the short bridge, as predicted in the DFT calculations. The amino species is basically atop, specifically the best fit placed the amino N atoms slightly off atop, but with a lateral displacement that is an order of magnitude less than the uncertainty; the imino species is instead found to be 0.60 ± 0.15 Å off atop. The full structural parameters for this best fit model can be found in Table 1 (cf. also Fig. 1c, d), along with the theoretical predictions, and the structure is drawn schematically in Fig. 5a, b. Importantly, models with the 2H-P molecule centered over the atop, hcp or fcc sites generally resulted in significantly higher R-factors (> 0.4), which were well beyond the variance.

**3.4 PhD analysis of Cu-P on Cu(111)**

The strongest modulations, for the Cu-P species, are shown in Fig. 4. Again, the most intense modulations are observed at normal emission, where they bear a remarkable similarity to that of the iminic N atoms of 2H-P. When modelling the data with multiple scattering calculations, the global searches indicated several numerically good fits ($R_{fac}$ = 0.28), however they resulted from structures with either the N-N distance being too short (< 3 Å) or too long (> 4.5 Å). Thus additional searches were performed, which were constrained around the structural parameters taken from the DFT study described above. The best R-factor found under these constraints was 0.33 (cf. Fig. 4), and indicated an average difference in height between the two sets of N species of ~0.35 Å. The N atoms astride the long bridge ($N_{long}$) axis are now closer to the Cu(111) surface (2.00 ± 0.02 Å) than the N atoms astride the short bridge ($N_{short}$): 2.30 / 2.40 ± 0.10 Å. The best fit placed the $N_{short}$ atoms further off atop than the imino species in the 2H-P molecule, but by an amount that is not significant compared to the uncertainties. The full structural parameters of this fit are reported in Table 1, and the resulting structure is depicted schematically in Fig. 5 c, d. Models, with the molecule rotated azimuthally by 90° resulted in significantly higher R-factors (> 0.4) exceeding the variance.

A second model was found with a similar R-factor of 0.31. Only three of the fitting parameters (listed in Table 1) differed significantly with above model: the adsorption height of $N_{short}$ (2.09 ± 0.07 Å), the relaxation of the Cu atoms under $N_{long}$ (0.1 ± 0.2 Å) and the lateral displacement of the $N_{short}$ atoms (0.3 ± 0.4 Å). Therefore this model has all four N atoms at, basically, the same adsorption height and the $N_{short}$ atoms far closer to atop, resulting in a dramatically smaller N-N distance of less than 3.5 Å. Such a short N-N distance is likely infeasible, as the determined molecular crystal structures of similar molecules, using X-ray diffraction, typically measure a distance of ~4.0-4.1 Å [82-84]. The occurrence of multiple minima in the multiple scattering analysis, e.g. for the R-factor as a function of the adsorption height, is in general not surprising [85, 86]. However, this second model yields much shorter N-Cu bond distances (by ~0.2 - 0.3 Å), which are neither supported by the DFT calculations (cf. Table 1) nor by the typical Cu-N bond distances in adsorbed systems (exceeding 1.95-2.00 Å [87-91]).

Regardless, a structure containing all four atoms at a similar adsorption height on the surface might possibly suggest that a mixed model, containing half of the molecules with $N_{long}$ closer to the surface, and half with $N_{long}$ further from the surface, than $N_{short}$, could provide good agreement with the experimental PhD modulations. This possibility is also inspired by the prediction of two energetically compatible structures essentially rotated by 90° for cobalt porphine[41], as discussed above. However, multiple-scattering calculations with 50%/50% mixing of these structures yielded significantly higher R-factors, providing no support for either such a mixed model or the occurrence of a 90° rotated molecular conformation. This finding is thus fully consistent with our DFT calculations for Cu-P detailed in section 3.1.

Finally, models with the molecule centered over atop, hcp or fcc sites were also tested, and also resulted in significantly higher R-factors.

## 4. Discussion

The agreement between theory and experiment for the 2H-P and Cu-P species is summarized in Table 1 for the respective optimal structure compatible with both the PhD and DFT analysis. The agreement is good, and, excluding the N-Cu bond distance of the amino (N-H) species in the 2H-P molecule, the structural parameters predicted by the DFT calculations are within two standard deviations of the experimental results. A global search of structural models for Cu-P, to compare against the PhD data, did return a second model that numerically fitted the experimental data well, however the N-N distance over the short-bridge direction is likely too short (< 3.5 Å) to be realistic, and does not agree with the model predicted by the DFT calculations.

The mean N adsorption heights measured for 2H-P, Cu-P, cobalt porphine[41], 2H-TPP[53] and Cu-TPP[53] are detailed in Table 2. Notably, there is a remarkable similarity between the adsorption heights of 2H-P (PhD) / 2H-TPP (NIXSW) and Cu-P (PhD) / Cu-TPP (NIXSW), with all heights being within the experiments' respective uncertainties. Our findings thus clearly suggest that substituted phenyl rings have surprisingly little influence on the surface anchoring of the central macrocycle through the N atoms. In the case of 2H-TPP/Cu(111) this might be related to the fact that the phenyl rings are not very much tilted from the surface plane, exhibiting an average tilt angle of 20°[46], unlike in 2H-TPP/Ag(111) where a tilt of 53° of the phenyl substituents was found[92].

Interestingly, combined STM and DFT studies by Lepper and co-workers[55, 80, 93] have recently proposed a so-called "inverted" adsorption structure of non-metalated porphyrins on Cu(111). In this model, the pyrrole groups of the imino species are tilted by almost 90° with respect to the flat-lying pyrrole groups of the amino species, in order to accommodate the substituted phenyl rings lying largely parallel with the underlying surface. This conformation allows strong covalent bonds to be formed between the iminic N atoms and the surface Cu atoms, as was observed here for 2H-P, and would result in the two iminic N atoms residing above bridge sites[80]. Such an "inverted" model is consistent with an earlier noncontact atomic force microscopy (AFM) investigation[94] that inferred a strong tilt of the iminic pyrrole rings, albeit without fully clarifying the structural details. Moreover, the "inverted" model could also be consistent with previous NEXAFS data on 2H-TPP/Cu(111)[46], which concluded a large 40° tilt averaged over the four pyrrole rings. In this regard, the apparent similarities between the N adsorption heights of 2H-P and 2H-TPP (see Table 2) may then suggest the prominent role of the iminic N-Cu covalent interaction in driving the "inversion" (possibly promoted by the presence of peripheral phenyl substituents), so as to ensure an optimal N-Cu covalent bond distance. Therefore, a similar PhD study, or a NIXSW triangulation study (such as that performed by Schwarz *et al.*[41]) could be decisive in resolving the issue, by proving quantitatively the existence of this "inverted" structure in a direct way and determining precisely the lateral registry of the nitrogen atoms on the surface. In turn, this would also provide deeper insight into the mechanism underlying the conformational change of the macrocycle from planar/saddle to "inverted".

In Table 2, the DFT-predicted values for the adsorption energy and charge transfer from the substrate into the molecules are also listed. Comparing the three related 2H-P, Cu-P and cobalt porphine[41] species, an increase in the adsorption strength is predicted by the DFT calculations through the molecules' progression. This trend, notably, anti-correlates the measured increase in adsorption heights from 2H-P to cobalt porphine through Cu-P. Note that cobalt porphine exhibits the highest average height of its N atoms, which is presumably due to the pronounced non-planarity of the macrocycle. As for the charge transfer, a minor increase from 2H-P to Cu-P is predicted by DFT, which might naively suggest a slight strengthening of the interaction with the substrate upon coordination of the Cu center; however, this is contrasted by a significant reduction of the charge transfer to cobalt porphine[41], despite the stronger adsorption energy of the latter. Thus, while the charge transfer and the distance between the adsorbate and the substrate undoubtedly play a role in how strongly materials interact with one another, our DFT calculations, benchmarked by quantitative experimental measurements, call into question generalized "rules of thumb" correlating the bond strength with either the adsorption height or the amount of static charge transfer. Such "rules of thumb" should therefore be treated with caution.

The availability of experimental techniques, such as PhD and NIXSW, enabling quantitative structure determination of organic adsorbates on single-crystal surfaces is of primary importance to benchmark DFT-predicted structural models. The DFT calculations play, on the other hand, a crucial role in rationalizing interfacial interactions and bonding strengths. This work, together with the previous study by Schwarz *et al.*[41], lends credence to the predictive power of DFT concerning structural models. Thus far, there is no direct quantification of metal substrate to molecule charge transfer, although qualitative evidence has been observed. Specifically, for 2H-P, Cu-P and cobalt porphine charge transfer into the molecules was concluded based on the presence of repulsive intermolecular interactions, the partial occupation of the lowest unoccupied molecular orbital(s) and the binding energy position of the metal ion core levels[37, 38, 41]. Such charge transfer can indeed be hampered by insertion of a decoupling layer of hexagonal boron nitride, as demonstrated in Refs. 39, 95. Interestingly, charge-transfer-induced electrostatic repulsive interactions were also shown for 2H-TPP on Cu(111)[96], preventing the formation of ordered two-dimensional networks in striking contrast to the Ag(111) surface, where attractive interactions govern the 2H-TPP self-assembly[96]. This might in principle support the previously discussed "inverted" porphyrins, where strong charge transfer is predicted by DFT in contrast to the saddle shape (i.e. "not inverted") counterpart. In any case, being that the theoretically predicted charge transfer values are strongly dependent on both the partial charge partitioning scheme and the molecular coverage[40], systematic coverage-dependent experimental studies to probe both occupied and unoccupied states, e.g. by means of NEXAFS and ultraviolet photoelectron spectroscopy or the more recently developed photoemission tomography[97], would be highly desirable. It has been shown that a semi-quantitative evaluation of fractional charge transfer at organic-metal interfaces can be, for example, achieved by comparing non-rigid shifts in core level photoemission with DFT calculations[98]. Similarly, quantification of adsorption energies of organic molecules on single-crystal surfaces would yield very valuable information to further test the theoretical analysis. For example, single-crystal adsorption calorimetry is an elegant method that enables the heat of adsorption to be measured[99-102].

## 5. Summary and Conclusions

In this work, the geometric structure of 2H-P and Cu-P on Cu(111) was determined experimentally by PhD and compared to dispersion-corrected DFT calculations. The lateral organization of Cu-P was also studied by means of LEED. Specifically, both molecules are found to adsorb with their centers above a surface bridge site (as displayed schematically in Fig. 5a-d), and the Cu-P molecules form a long-range ordered overlayer with $p(3\sqrt{3} \times 3\sqrt{3})R30°$ periodicity (see Fig. 5e). The measured adsorption heights of the 2H-P and Cu-P molecules agree very well with the results of a NIXSW study by Bürker et al.[53], suggesting that the addition of peripheral phenyl rings to the porphine macrocycle has limited impact on the bonding between the N atoms and the copper surface. While substituted phenyl rings hardly affect the average adsorption height of the N atoms of the porphyrin macrocycle, the character of the central cavity of the porphine molecule plays a significant role. After incorporation of a copper atom into the macrocycle, the N atoms astride the long bridge direction are closer to the Cu(111) surface than the N atoms astride the perpendicular short bridge direction, whereas the opposite behavior is found and predicted for the free-base 2H-P molecule. As a result, the mean adsorption height of the N atoms is higher for Cu-P than for 2H-P. The combined PhD structural analysis and DFT calculations, along with the comparison with an analogous study of cobalt porphine on the same Cu(111) surface, indicate that caution should be taken in using the adsorption height as sole indicator of interaction strengths, whereby also the amount of charge transfer cannot be univocally correlated with the adsorption strength.

The in-depth comparison of the determined adsorption heights of 2H-P, Cu-P and cobalt porphine[41] shows the distinct structural difference induced by the specific character of the center of the macrocycle. Specifically, the mean adsorption height of the central N atoms increases by 0.06 ± 0.11 Å (Cu-P) and 0.19 ± 0.0 Å (cobalt porphine), respectively, with respect to 2H-P. The character of the center of the macrocycle is known to have a significant effect on the chemistry of the molecule, notably, in the tetraphenyl porphyrins, dictating the selectivity of intra-molecular reactions on the periphery of the molecule[103], however here we clearly demonstrate a corresponding structural difference. What is more surprising is that such differences do not manifest from introducing substituent groups on the periphery of the molecule. Indeed, with tetraphenyl substituents in the *meso*-position, the mean adsorption height of the central nitrogen atoms does not differ, within the uncertainty of the experimental techniques, between 2H-P and 2H-TPP (or Cu-P and Cu-TPP)[53]. As recent studies[55, 80, 93] indicate a significant distortion of the macrocycle upon adsorption of 2H-TPP on Cu(111), this lack of difference in the mean adsorption height of the N atoms is highly surprising, and would support the conclusion that, if the macrocycle is adsorbed with such a molecular distortion, this distortion is driven by the strong interaction between the N atoms and the Cu substrate. Conversely it would also suggest that, although the character of the center of the macrocycle plays a role in the chemistry of the periphery of the porphyrin molecule, the influence of the peripheral, substituent groups on the chemistry of the center of the macrocycle may be far more muted.

Finally, our study shows that combined PhD and DFT investigations are valuable in providing quantitative structural description of organic/metal interfaces. Though PhD can result in multiple models that numerically fit with the experimental data, DFT can exclude those that are physically improbable; conversely, DFT may sometimes predict multiple adsorption models that are energetically compatible, and PhD can exclude those that are simply not consistent with the fit.

**Supplementary Data**

The DFT-predicted structures for 2H-P/Cu(111) and Cu-P/Cu(111) are provided as supplementary data in XYZ file format.

**Acknowledgements**

This work was supported by the Munich-Centre for Advanced Photonics (MAP), the ERC Advanced Grants MolArt (Project ID 247299) and QSpec-NewMat (Project ID 694097), the German Research Foundation (DFG) *via* KL 2294/3-1, and Grupos Consolidados UPV/EHU (IT578-13). D.A.D. acknowledges funding from the Alexander von Humboldt Foundation and the Marie Curie Intra-European Fellowship for Career Development (SiliNano, project no. 626397). The authors gratefully acknowledge the computing resources provided by the Leibniz Supercomputing Centre *(*www.lrz.de*)* and the STFC Scientific Computing Department's SCARF cluster. We would also like to thank Elettra Sincrotrone Trieste for the award of beam time and Willi Auwärter for useful discussion.

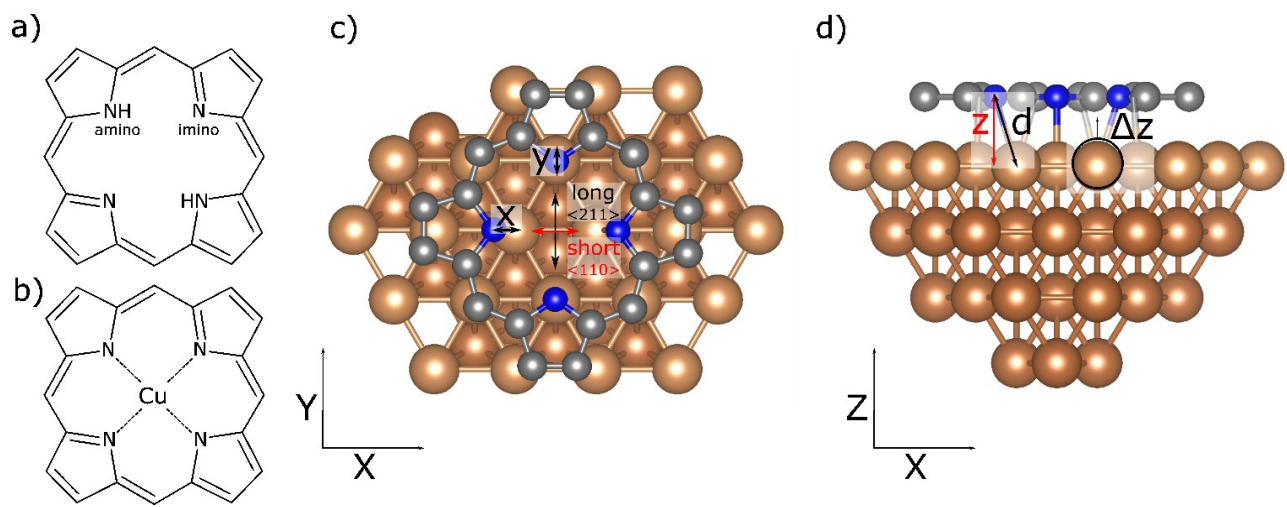

**Fig. 1:** Molecular structure of (a) 2H-P and (b) Cu-P. A schematic of the porphine molecule above a bridge site on the surface is shown in (c) top and (d) side view. The long and short bridge directions are indicated, along with the relevant $x$, $y$, $z$ and $d$ parameters that are determined in the PhD and DFT structural analysis (cf. Table 1). The grey, blue, and copper colored spheres represent the carbon, nitrogen and copper atoms, respectively. H atoms and copper centers are omitted for clarity. Note that the actual bending and conformation of the macrocycle are not rendered in the sketches of (c) and (d).

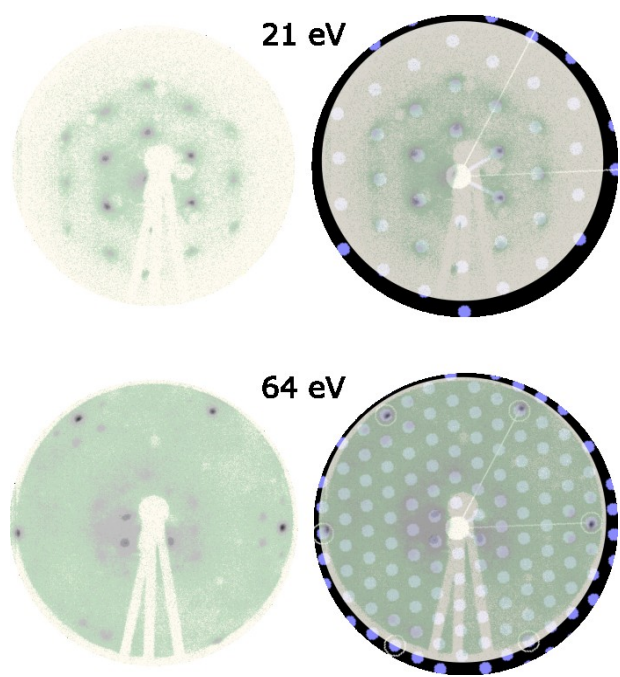

**Fig. 2:** Measured LEED pattern from a full (annealed) layer of Cu-P on Cu(111) (left) and the same pattern with the $p(3\sqrt{3} \times 3\sqrt{3})R30°$ superstructure overlaid (right). Primary electron energy: 21 eV (top) and 64 eV (bottom).

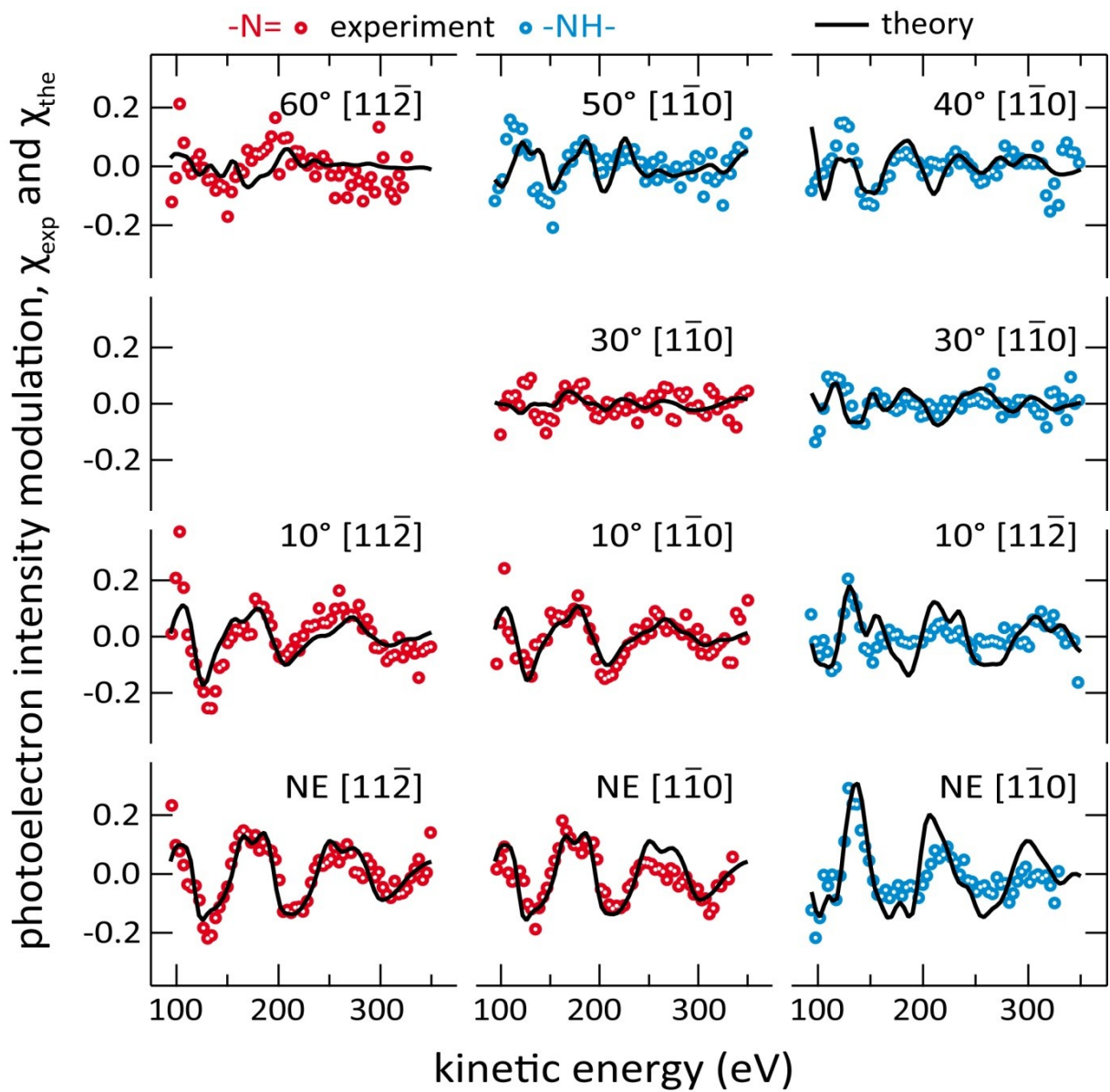

**Fig. 3**: N 1s PhD data from a full layer of 2H-P on Cu(111) compared against the best fitting model structure, $R_{fac}$ = 0.34, shown in Fig. 5a, b. All polar emission angles are specified with respect to the surface normal (normal emission, NE: 0°).

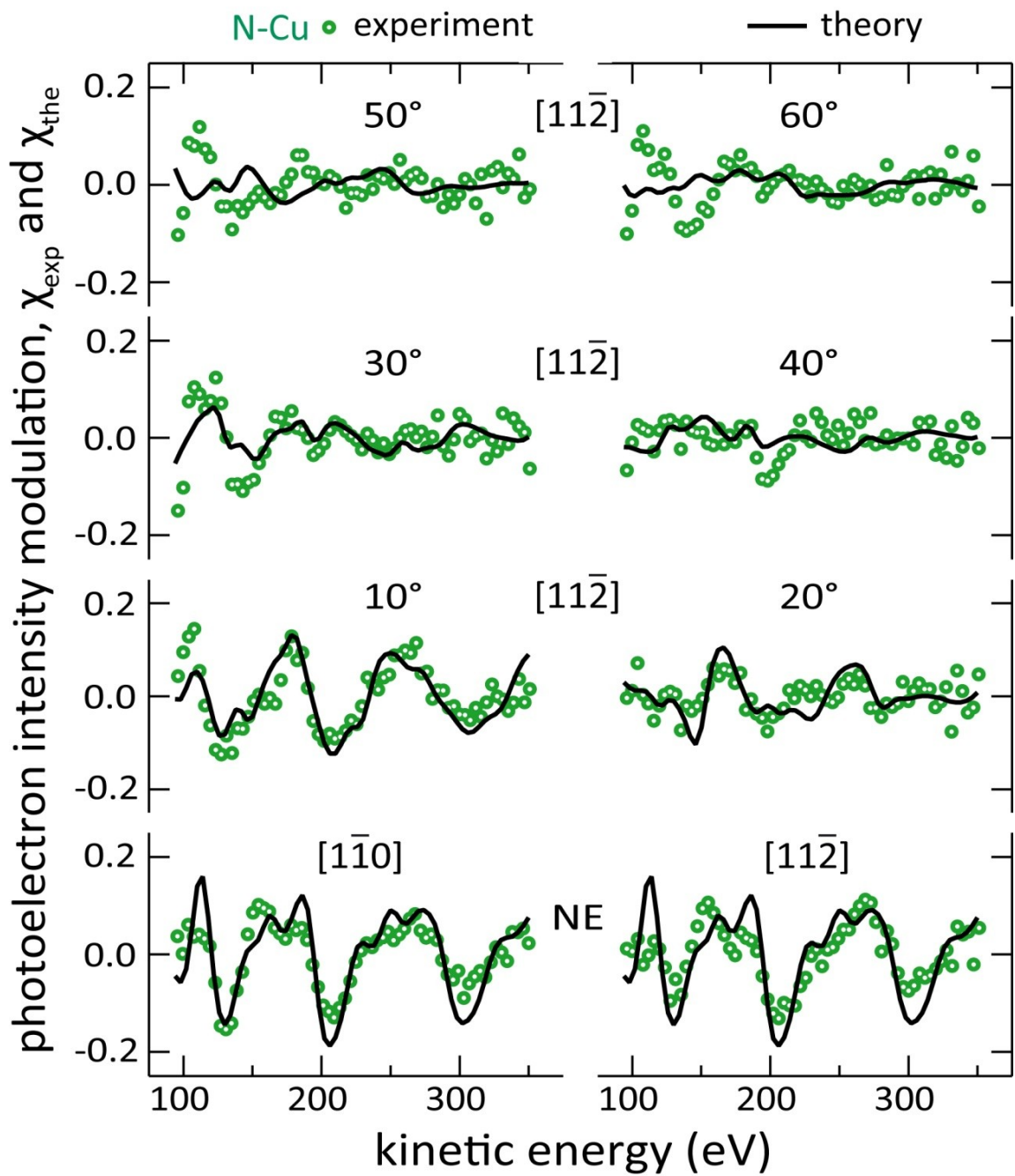

**Fig. 4**: N 1$s$ PhD data from a full layer of Cu-P on Cu(111) compared against the best fitting model structure, R$_{fac}$ = 0.33, shown in Fig. 5c, d. All polar emission angles are specified with respect to the surface normal (normal emission, NE: 0°).

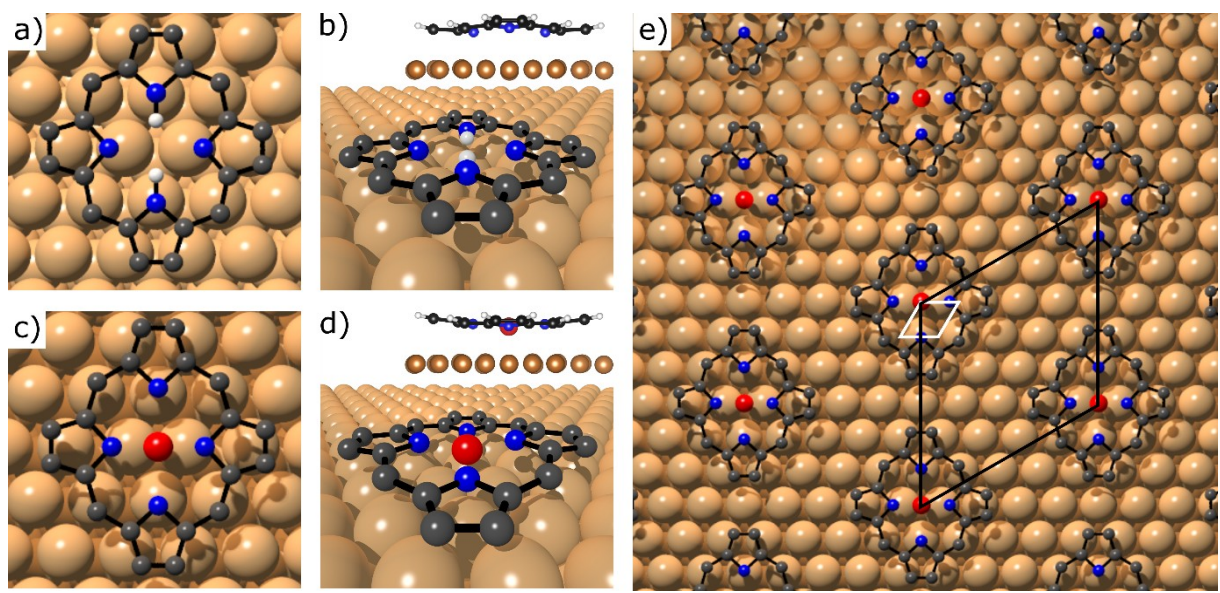

**Fig. 5:** Model of the best fitting structure for (a, b) 2H-P and (c, d) Cu-P adsorbed on Cu(111) shown from (a, c) above and from (b, d) the side, as determined by the PhD analysis. Inset in (b, d) is the corresponding side view of the predicted DFT structure. (e) Proposed lateral structure of Cu-P on Cu(111) based on the PhD and LEED measurements (Fig. 2). The unit cell of the $p(3\sqrt{3} \times 3\sqrt{3})R30°$ Cu-P overlayer and Cu(111) surface are drawn in black and white, respectively. The H, N, C, surface Cu, and central Cu atoms are depicted by the white, blue, dark grey, light brown, and red spheres, respectively. Note that H atoms are omitted from the experimental structures (they are not modelled in the multiple scattering PhD calculations), except the two H atoms (white spheres in (a) and (b)) of the amino groups in 2H-P, which are solely shown to differentiate the inequivalent N atoms in 2H-P.

**Table 1:** Structural parameters predicted by theoretical calculations (DFT), compared to those from the PhD-derived best fitting model structures of Figs. 3 and 4. $Cu_{short}$ and $Cu_{long}$ are the nearest neighbor Cu atoms astride the short and long bridge site directions (respectively), indicated schematically in Fig. 1c. The label *d* refers to the bond length of the N atoms to the nearest Cu atoms underneath, whereas the coordinate *z* denotes the vertical height of the respective N atom with respect to the average Cu(111) top layer. Coordinates *x* and *y* are absolute values of the lateral displacements of the N atoms with respect to the unrelaxed positions of the nearest Cu atom underneath, with *x* aligned along the close packed [1$\bar{1}$0] rows; the displacements are symmetric for the two atoms of the same species ($N_{short}$ or $N_{long}$), with both atoms moving away the central bridge site along the respective axis (short bridge or long bridge). The quantities Δ*z* describe, respectively, the vertical displacements of the substrate Cu atoms directly below the N atoms of the molecules ($Cu_{short}$, $Cu_{long}$) relative to the mean height of the top Cu(111) layer, and the relaxation of the first copper layer (Δ*z* 1$^{st}$ layer Cu) relative to a bulk terminated surface. Finally, *z* Cu center defines the height of the central Cu ion of Cu-P relative to the average N plane. The structural parameters are explained in Fig. 1c, d. Note that the adsorbed 2H-P molecule has the amino (imino) species aligned with the long (short) bridge direction.

| 2H-P / Cu-P | 2H-P DFT (Å) | 2H-P Experiment (Å) | Cu-P DFT (Å) | Cu-P Experiment (Å) |
|---|---|---|---|---|
| *d* imino / $N_{short}$ - Cu | 2.15 | 2.11 ± 0.05 | 2.34 | 2.40/2.50 ± 0.10 |
| *d* amino / $N_{long}$ - Cu | 2.43 | 2.27 ± 0.05 | 2.14 † | 2.11 ± 0.03 † |
| *z* imino / $N_{short}$ | 2.09 | 2.08 ± 0.03 | 2.21 | 2.30/2.40 ± 0.10 |
| *z* amino / $N_{long}$ | 2.33 | 2.20 ± 0.05 | 2.10 † | 2.00 ± 0.02 † |
| *x* imino / $N_{short}$ | 0.68 | 0.60 ± 0.15 | 0.59 | 0.7 (-0.4/+0.1) |
| *y* amino / $N_{long}$ | 0.01 | 0.0 ± 0.4 | 0.12 † | 0.2 (-0.3/+0.2) † |
| Δ*z* $Cu_{short}$ | 0.05 | 0.05 (-0.10/+0.20) | -0.06 | -0.0 (-0.1/+*) |
| Δ*z* $Cu_{long}$ | -0.10 | -0.07 ± 0.10 | -0.04 | -0.10 ± 0.07 |
| Δ*z* 1$^{st}$ layer Cu | -0.14 | -0.05 (-0.15/+*) | -0.13 | -0.05 ± 0.08 |
| *z* Cu center | -- | -- | -0.09 | -0.1 (-0.7/+*) |

*† Note: Cu-P, of course, does not have amino groups, but these N atoms occupy the same lateral adsorption site as the amino species in 2H-P.*

*\* The R-factor is insensitive to variations of the parameter in this direction.*

**Table 2:** Comparison of the experimental mean N adsorption heights and the DFT-predicted charge transfer and adsorption energy found for cobalt porphine[41], 2H-TPP[53, 80], Cu-TPP[53, 80], 2H-P and Cu-P on the Cu(111) surface. All data refer to measurements performed at RT.

|  | cobalt porphine Ref. 41 | 2H-TPP | Cu-TPP Ref. 53 | 2H-P | Cu-P |
|---|---|---|---|---|---|
| $z$ N$_{mean}$ (exp. / Å) | 2.33 ± 0.06 | 2.15 ± 0.08 Ref. 53 | 2.25 ± 0.02 | 2.14 ± 0.05 | 2.20 ± 0.10 |
| q $_{Cu(111) \rightarrow molec.}$ (DFT / e$^-$) | 0.66 | 0.08 / 1.91*† Ref. 80 | -- | 1.08 | 1.26 |
| E$_{ads}$ (DFT / eV) | -4.797 | -6.78 / -7.13*‡ Ref. 80 | -- | -4.032 | -4.257 |

*for the saddle / "inverted" models.
† calculated using PBE-D3[104].
‡ calculated using the Tkatchenko-Scheffler approach[74], similar to the work presented here.

**References:**

[1] L. Pauling, *The Nature of the Chemical Bond* (Ithaca, N.Y Cornell University Press, 1960), 3rd edn.,
[2] I. D. Brown, and R. D. Shannon, Acta Cryst. A **A29**, 266 (1973).
[3] J. Zió łkowski, and L. Dziembaj, J. Solid State Chem. **57**, 291 (1985).
[4] P. Politzer, and S. Ranganathan, Chem. Phys. Lett. **124**, 527 (1986).
[5] G. V. Gibbs, F. C. Hill, M. B. Boisen, and R. T. Downs, Phys. Chem. Miner. **25**, 585 (1998).
[6] M. Kaupp, B. Metz, and H. Stoll, Angew. Chem. Int. Ed. **39**, 4607 (2000).
[7] A. Nimmermark, L. Öhrström, and J. Reedijk, Z. Kristallogr. **228**, 311 (2013).
[8] B. A. Lindquist, and T. H. Dunning, J. Phys. Chem. Lett. **4**, 3139 (2013).
[9] M. Kaupp, D. Danovich, and S. Shaik, Coord. Chem. Rev. **344**, 355 (2017).
[10] A. Haaland, Angew. Chem. Int. Ed. **28**, 992 (1989).
[11] D. I. Sayago, J. T. Hoeft, M. Polcik, M. Kittel, R. L. Toomes, J. Robinson, D. P. Woodruff, M. Pascal, C. L. A. Lamont, and G. Nisbet, Phys. Rev. Lett. **90**, 116104 (2003).
[12] B. Stadtmüller, D. Lüftner, M. Willenbockel, E. M. Reinisch, T. Sueyoshi, G. Koller, S. Soubatch, M. G. Ramsey, P. Puschnig, and F. S. Tautz, Nat. Commun. **5**, 3685 (2014).
[13] A. Franco-Cañellas, Q. Wang, K. Broch, D. A. Duncan, P. K. Thakur, L. Liu, S. Kera, A. Gerlach, S. Duhm, and F. Schreiber, Phys. Rev. Mater. **1**, 013001 (2017).
[14] S. Duhm, A. Gerlach, I. Salzmann, B. Bröker, R. L. Johnson, F. Schreiber, and N. Koch, Org. Electron. **9**, 111 (2008).
[15] B. Hulsken, R. Van Hameren, J. W. Gerritsen, T. Khoury, P. Thordarson, M. J. Crossley, A. E. Rowan, R. J. M. Nolte, J. A. A. W. Elemans, and S. Speller, Nat. Nanotechnol. **2**, 285 (2007).
[16] B. E. Murphy, S. A. Krasnikov, N. N. Sergeeva, A. A. Cafolla, A. B. Preobrajenski, A. N. Chaika, O. Lübben, and I. V. Shvets, ACS Nano **8**, 5190 (2014).

[17] K. Seufert, W. Auwärter, and J. V. Barth, J. Am. Chem. Soc. **132**, 18141 (2010).

[18] A. Yella, H.-W. Lee, H. N. Tsao, C. Yi, A. K. Chandiran, M. K. Nazeeruddin, E. W.-G. Diau, C.-Y. Yeh, S. M. Zakeeruddin, and M. Grätzel, Science **334**, 629 (2011).

[19] L. Bogani, and W. Wernsdorfer, Nat. Mater. **7**, 179 (2008).

[20] Z. Xiong, D. Wu, Z. V. Vardeny, and J. Shi, Nature **427**, 821 (2004).

[21] X. Zhang, Y. Shen, S. Wang, Y. Guo, K. Deng, C. Wang, and Q. Zeng, Sci. Rep. **2**, 742 (2012).

[22] W. Auwärter, D. Écija, F. Klappenberger, and J. V. Barth, Nat. Chem. **7**, 105 (2015).

[23] J. M. Gottfried, Surf. Sci. Rep. **70**, 259 (2015).

[24] T. Bathon, P. Sessi, K. A. Kokh, O. E. Tereshchenko, and M. Bode, Nano Lett. **15**, 2442 (2015).

[25] L. Cui, Y.-F. Geng, C. F. Leong, Q. Ma, D. M. D'Alessandro, K. Deng, Q.-D. Zeng, and J.-L. Zuo, Sci. Rep. **6**, 25544 (2016).

[26] V. Q. Nguyen, X. Sun, F. Lafolet, J.-F. Audibert, F. Miomandre, G. Lemercier, F. Loiseau, and J.-C. Lacroix, J. Am. Chem. Soc. **138**, 9381 (2016).

[27] F. Klappenberger, Prog. Surf. Sci. **89**, 1 (2014).

[28] K. Diller, A. C. Papageorgiou, F. Klappenberger, F. Allegretti, J. V. Barth, and W. Auwärter, Chem. Soc. Rev. **45**, 1629 (2016).

[29] A. Wiengarten, K. Seufert, W. Auwärter, D. Écija, K. Diller, F. Allegretti, F. Bischoff, S. Fischer, D. A. Duncan, A. C. Papageorgiou, F. Klappenberger, R. G. Acres, T. H. Ngo, and J. V. Barth, J. Am. Chem. Soc. **136**, 9346 (2014).

[30] Y. He, M. Garnica, F. Bischoff, J. Ducke, M.-L. Bocquet, M. Batzill, W. Auwärter, and J. V. Barth, Nat. Chem. **9**, 33 (2017).

[31] P. S. Deimel, R. M. Bababrik, B. Wang, P. J. Blowey, L. A. Rochford, P. K. Thakur, T.-L. Lee, M.-L. Bocquet, J. V. Barth, D. P. Woodruff, D. A. Duncan, and F. Allegretti, Chem. Sci. **7**, 5647 (2016).

[32] W. Hieringer, K. Flechtner, A. Kretschmann, K. Seufert, W. Auwärter, J. V. Barth, A. Görling, H.-P. Steinrück, and J. M. Gottfried, J. Am. Chem. Soc. **133**, 6206 (2011).

[33] C. Stadler, S. Hansen, I. Kröger, C. Kumpf, and E. Umbach, Nat. Phys. **5**, 153 (2009).

[34] J. Beggan, S. Krasnikov, N. Sergeeva, M. Senge, and A. Cafolla, J. Phys.: Condens. Matter **20**, 015003 (2007).

[35] S. A. Krasnikov, J. P. Beggan, N. N. Sergeeva, M. O. Senge, and A. A. Cafolla, Nanotechnol. **20**, 135301 (2009).

[36] M. S. Dyer, A. Robin, S. Haq, R. Raval, M. Persson, and J. i. Klimeš, ACS Nano **5**, 1831 (2011).

[37] F. Bischoff, K. Seufert, W. Auwärter, S. Joshi, S. Vijayaraghavan, D. Écija, K. Diller, A. C. Papageorgiou, S. Fischer, and F. Allegretti, ACS Nano **7**, 3139 (2013).

[38] K. Diller, F. Klappenberger, F. Allegretti, A. Papageorgiou, S. Fischer, A. Wiengarten, S. Joshi, K. Seufert, D. Écija, and W. Auwärter, J. Chem. Phys. **138**, 154710 (2013).

[39] S. Joshi, F. Bischoff, R. Koitz, D. Écija, K. Seufert, A. P. Seitsonen, J. r. Hutter, K. Diller, J. I. Urgel, and H. Sachdev, ACS Nano **8**, 430 (2013).

[40] M. Müller, K. Diller, R. J. Maurer, and K. Reuter, J. Chem. Phys. **144**, 024701 (2016).

[41] M. Schwarz, M. Garnica, D. A. Duncan, A. Pérez Paz, J. Ducke, P. S. Deimel, P. K. Thakur, T.-L. Lee, A. Rubio, J. V. Barth, F. Allegretti, and W. Auwärter, J. Phys. Chem. C **122**, 5452 (2018).

[42] H. Marbach, Acc. Chem. Res. **48**, 2649 (2015).

[43] D. A. Duncan, P. S. Deimel, A. Wiengarten, R. Han, R. G. Acres, W. Auwärter, P. Feulner, A. C. Papageorgiou, F. Allegretti, and J. V. Barth, Chem. Commun. **51**, 9483 (2015).

[44] T. E. Shubina, H. Marbach, K. Flechtner, A. Kretschmann, N. Jux, F. Buchner, H.-P. Steinrück, T. Clark, and J. M. Gottfried, J. Am. Chem. Soc. **129**, 9476 (2007).

[45] A. C. Papageorgiou, S. Fischer, S. C. Oh, O. Saglam, J. Reichert, A. Wiengarten, K. Seufert, S. Vijayaraghavan, D. Écija, and W. Auwärter, ACS Nano **7**, 4520 (2013).

[46] K. Diller, F. Klappenberger, M. Marschall, K. Hermann, A. Nefedov, C. Wöll, and J. Barth, J. Chem. Phys. **136**, 014705 (2012).

[47] A. Goldoni, C. A. Pignedoli, G. Di Santo, C. Castellarin-Cudia, E. Magnano, F. Bondino, A. Verdini, and D. Passerone, ACS Nano **6**, 10800 (2012).

[48] J. Schneider, M. Franke, M. Gurrath, M. Röckert, T. Berger, J. Bernardi, B. Meyer, H. P. Steinrück, O. Lytken, and O. Diwald, Chem. Eur. J. **22**, 1744 (2016).

[49] G. Lovat, D. Forrer, M. Abadia, M. Dominguez, M. Casarin, C. Rogero, A. Vittadini, and L. Floreano, J. Phys. Chem. C **121**, 13738 (2017).

[50] M. Röckert, S. Ditze, M. Stark, J. Xiao, H.-P. Steinrück, H. Marbach, and O. Lytken, J. Phys. Chem. C **118**, 1661 (2014).

[51] J. Xiao, S. Ditze, M. Chen, F. Buchner, M. Stark, M. Drost, H.-P. Steinrück, J. M. Gottfried, and H. Marbach, J. Phys. Chem. C **116**, 12275 (2012).

[52] S. Ditze, M. Stark, M. Drost, F. Buchner, H. P. Steinrück, and H. Marbach, Angew. Chem. Int. Ed. **51**, 10898 (2012).

[53] C. Bürker, A. Franco-Cañellas, K. Broch, T.-L. Lee, A. Gerlach, and F. Schreiber, J. Phys. Chem. C **118**, 13659 (2014).

[54] M. Stark, S. Ditze, M. Lepper, L. Zhang, H. Schlott, F. Buchner, M. Röckert, M. Chen, O. Lytken, and H.-P. Steinrück, Chem. Commun. **50**, 10225 (2014).

[55] M. Lepper, J. Köbl, L. Zhang, M. Meusel, H. Hölzel, D. Lungerich, N. Jux, A. de Siervo, B. Meyer, H.-P. Steinrück, and H. Marbach, Angew. Chem. Int. Ed. **57**, 10074 (2018).

[56] D. P. Woodruff, Surf. Sci. Rep. **62**, 1 (2007).

[57] D. P. Woodruff, and A. M. Bradshaw, Rep. Progr. Phys, **57**, 1029 (1994).

[58] P. Hofmann, K. M. Schindler, S. Bao, A. M. Bradshaw, and D. P. Woodruff, Nature **368**, 131 (1994).

[59] G. Drera, G. Salvinelli, J. Åhlund, P. G. Karlsson, B. Wannberg, E. Magnano, S. Nappini, and L. Sangaletti, J. Electron Spectrosc. Relat. Phenom. **195**, 109 (2014).

[60] D. K. Lorenzo, M. Bradley, W. Unterberger, D. Duncan, T. Lerotholi, J. Robinson, and D. Woodruff, Surf. Sci. **605**, 193 (2011).

[61] V. Fritzsche, Surf. Sci. **213**, 648 (1989).

[62] V. Fritzsche, J. Phys.: Condens. Matter **2**, 1413 (1990).

[63] V. Fritzsche, Surf. Sci. **265**, 187 (1992).

[64] V. Fritzsche, and J. B. Pendry, Phys. Rev. B **48**, 9054 (1993).

[65] V. Fritzsche, R. Davis, X. M. Hu, D. P. Woodruff, K. U. Weiss, R. Dippel, K. M. Schindler, P. Hofmann, and A. M. Bradshaw, Phys. Rev. B **49**, 7729 (1994).

[66] D. A. Duncan, J. I. J. Choi, and D. P. Woodruff, Surf. Sci. **606**, 278 (2012).

[67] J. B. Pendry, J. Phys.: Solid State Phys. **13**, 937 (1980).

[68] N. A. Booth, R. Davis, R. Toomes, D. P. Woodruff, C. Hirschmugl, K. M. Schindler, O. Schaff, V. Fernandez, A. Theobald, P. Hofmann, R. Lindsay, T. Gießel, P. Baumgärtel, and A. M. Bradshaw, Surf. Sci. **387**, 152 (1997).

[69] P. E. Blöchl, Phys. Rev. B **50**, 17953 (1994).

[70] G. Kresse, and J. Hafner, Phys. Rev. B **47**, 558 (1993).

[71] G. Kresse, and J. Hafner, J. Phys.: Condens. Matter **6**, 8245 (1994).

[72] G. Kresse, and J. Furthmüller, Phys. Rev. B **54**, 11169 (1996).

[73] J. P. Perdew, K. Burke, and M. Ernzerhof, Phys. Rev. Lett. **77**, 3865 (1996).

[74] A. Tkatchenko, and M. Scheffler, Phys. Rev. Lett. **102**, 073005 (2009).

[75] R. Koitz, A. P. Seitsonen, M. Iannuzzi, and J. Hutter, Nanoscale **5**, 5589 (2013).

[76] R. Cuadrado, J. I. Cerdá, Y. Wang, G. Xin, R. Berndt, and H. Tang, J. Chem. Phys. **133**, 154701 (2010).

[77] G. Henkelman, A. Arnaldsson, and H. Jónsson, Comput. Mater. Sci. **36**, 354 (2006).

[78] E. Sanville, S. D. Kenny, R. Smith, and G. Henkelman, J. Comput. Chem. **28**, 899 (2007).

[79] W. Tang, E. Sanville, and G. Henkelman, J. Phys.: Condens. Matter **21**, 084204 (2009).

[80] M. Lepper, J. Köbl, T. Schmitt, M. Gurrath, A. de Siervo, M. A. Schneider, H.-P. Steinrück, B. Meyer, H. Marbach, and W. Hieringer, Chem. Commun. **53**, 8207 (2017).

[81] M. A. van Hove, and K. E. Hermann, Berlin / Hong Kong, 2014).

[82] A. J. Golder, K. B. Nolan, D. C. Povey, and L. R. Milgrom, Acta Cryst. C **44**, 1916 (1988).

[83] L. E. Webb, and E. B. Fleischer, J. Chem. Phys. **43**, 3100 (1965).

[84] H.-H. Wang, W.-H. Wen, H.-B. Zou, F. Cheng, A. Ali, L. Shi, H.-Y. Liu, and C.-K. Chang, New J. Chem. **41**, 3508 (2017).

[85] S. D. Ruebush, R. E. Couch, S. Thevuthasan, and C. S. Fadley, Surf. Sci. **421**, 205 (1999).

[86] F. Allegretti, S. O'Brien, M. Polcik, D. I. Sayago, and D. P. Woodruff, Surf. Sci. **600**, 1487 (2006).

[87] N. A. Booth, D. P. Woodruff, O. Schaff, T. Gießel, R. Lindsay, P. Baumgärtel, and A. M. Bradshaw, Surf. Sci. **397**, 258 (1998).

[88] T. Gießel, O. Schaff, R. Lindsay, P. Baumgärtel, M. Polcik, A. M. Bradshaw, A. Koebbel, T. McCabe, M. Bridge, D. R. Lloyd, and D. P. Woodruff, J. Chem. Phys. **110**, 9666 (1999).

[89] F. Allegretti, M. Polcik, and D. P. Woodruff, Surf. Sci. **601**, 3611 (2007).

[90] D. A. Duncan, W. Unterberger, D. Kreikemeyer-Lorenzo, and D. P. Woodruff, J. Chem. Phys. **135**, 014704 (2011).

[91] D. A. Duncan, M. K. Bradley, W. Unterberger, D. Kreikemeyer-Lorenzo, T. J. Lerotholi, J. Robinson, and D. P. Woodruff, J. Phys. Chem. C **116**, 9985 (2012).

[92] W. Auwärter, K. Seufert, F. Bischoff, D. Écija, S. Vijayaraghavan, S. Joshi, F. Klappenberger, N. Samudrala, and J. V. Barth, Nat. Nanotechnol. **7**, 41 (2012).

[93] M. Lepper, T. Schmitt, M. Gurrath, M. Raschmann, L. Zhang, M. Stark, H. Hölzel, N. Jux, B. Meyer, M. A. Schneider, H.-P. Steinrück, and H. Marbach, J. Phys. Chem. C **121**, 26361 (2017).

[94] F. Albrecht, F. Bischoff, W. Auwärter, J. V. Barth, and J. Repp, Nano Lett. **16**, 7703 (2016).

[95] M. Schwarz, D. A. Duncan, M. Garnica, J. Ducke, P. S. Deimel, P. K. Thakur, T.-L. Lee, F. Allegretti, and W. Auwärter, Nanoscale **10**, 21971 (2018).

[96] G. Rojas, X. Chen, C. Bravo, J.-H. Kim, J.-S. Kim, J. Xiao, P. A. Dowben, Y. Gao, X. C. Zeng, W. Choe, and A. Enders, J. Phys. Chem. C **114**, 9408 (2010).

[97] G. Zamborlini, D. Lüftner, Z. Feng, B. Kollmann, P. Puschnig, C. Dri, M. Panighel, G. Di Santo, A. Goldoni, G. Comelli, M. Jugovac, V. Feyer, and C. M. Schneider, Nat. Commun. **8**, 335 (2017).

[98] S.-A. Savu, G. Biddau, L. Pardini, R. Bula, H. F. Bettinger, C. Draxl, T. Chassé, and M. B. Casu, J. Phys. Chem. C **119**, 12538 (2015).

[99] J. M. Gottfried, E. K. Vestergaard, P. Bera, and C. T. Campbell, J. Phys. Chem. B **110**, 17539 (2006).

[100] O. Lytken, H.-J. Drescher, R. Kose, and J. M. Gottfried, in *Surface Science Techniques*, edited by G. Bracco, and B. Holst (Springer Berlin Heidelberg, Berlin, Heidelberg, 2013), pp. 35.

[101] T. L. Silbaugh, and C. T. Campbell, J. Phys. Chem. C **120**, 25161 (2016).

[102] S. J. Carey, W. Zhao, Z. Mao, and C. T. Campbell, J. Phys. Chem. C, (2018).

[103] A. Wiengarten, J. A. Lloyd, K. Seufert, J. Reichert, W. Auwärter, R. Han, D. A. Duncan, F. Allegretti, S. Fischer, S. C. Oh, Ö. Sağlam, L. Jiang, S. Vijayaraghavan, D. Écija, A. C. Papageorgiou, and J. V. Barth, Chem. Eur. J. **21**, 12285 (2015).

[104] S. Grimme, J. Antony, S. Ehrlich, and H. Krieg, J. Chem. Phys. **132**, 154104 (2010).